\documentclass[fleqn,usenatbib]{mnras}

\usepackage{newtxtext,newtxmath}
\usepackage{color}
\usepackage[T1]{fontenc}
\usepackage[symbol]{footmisc}
\DeclareRobustCommand{\VAN}[3]{#2}
\let\VANthebibliography\thebibliography
\def\thebibliography{\DeclareRobustCommand{\VAN}[3]{##3}\VANthebibliography}

\usepackage{graphicx}	
\usepackage{amsmath}	
\usepackage[normalem]{ulem}
\usepackage{lineno}

\title[BH and relativistic jet formation]{Numerical relativity simulations of black hole and relativistic jet formation}

\author[T. Kuroda and M. Shibata]{
Takami Kuroda$^{1}$\thanks{E-mail: takami.kuroda@aei.mpg.de}
and Masaru Shibata$^{1,2}$
\\
$^{1}$Max-Planck-Institut f{\"u}r Gravitationsphysik, Am M{\"u}hlenberg 1, D-14476 Potsdam-Golm, Germany\\
$^{2}$Center for Gravitational Physics and Quantum-Information,
Yukawa Institute for Theoretical Physics, Kyoto University, Kyoto, 606-8502, Japan
}

\date{Accepted XXX. Received YYY; in original form ZZZ}

\pubyear{2024}

\begin{document}
\label{firstpage}
\pagerange{\pageref{firstpage}--\pageref{lastpage}}
\maketitle

\begin{abstract}
We investigate impacts of stellar rotation and magnetic fields on black hole (BH) formation and its subsequent explosive activities, by conducting axisymmetric radiation-magnetohydrodynamics simulations of gravitational collapse of a $70$\,$M_\odot$ star with two-moment multi energy neutrino transport in full general relativity for the first time.
Due to its dense stellar structure, all models cannot avoid the eventual BH formation even though a strongly magnetized model experiences the so-called magnetorotational explosion prior to the BH formation.
One intriguing phenomenon observed in the strongly magnetized model is the formation of a relativistic jet in the post-BH formation.
The relativistic jet is the outcome of a combination of strong magnetic fields and low-density materials above the BH.
The jet further enhances the explosion energy beyond $\sim10^{52}$\,erg, which is well exceeding the gravitational overburden ahead of the shock.
Our self-consistent supernova models demonstrate that rotating magnetized massive stars at the high-mass end of supernova progenitors could be a potential candidate of hypernova and long gamma-ray burst progenitors.
\end{abstract}




\maketitle

\section{Introduction}
\label{sec:Introduction}
Core collapse of magnetized rotating massive stars and their subsequent magnetohydrodynamic activities in the vicinity of a spinning black hole (BH) may be the key to understanding a subclass of supernovae (SNe), which exhibit extra ordinary explosion energies, referred to often as hypernovae (HNe), and are sometimes accompanied by a long gamma-ray burst \citep[LGRB:][]{Woosley93}.

Massive stars heavier than about $8\,M_\odot$ terminate their evolution with a gravitational core collapse and leave behind various compact stars including BH \citep{WHW02,Heger03}.
Although the progenitor mass dependence of the BH formation is still ambiguous, many of previous SN simulations have reported that BHs are likely born from collapse of the high-mass end of SN progenitors, whose zero-age-main-sequence (ZAMS) mass $M_{\rm ZAMS}$ is $M_{\rm ZAMS}\sim30$--$40$\,$M_\odot$ or heavier (see, e.g., \cite{Liebendorfer04,Sumiyoshi06,Fischer09,O'Connor11} for one-dimensional (1D) spherical symmetry studies and \cite{Walk20,Rahman2022,Shibagaki23,KurodaT23BH} for recent multi-D ones).
Such progenitor mass dependence on the BH formation might be also related to the observed lower mass gap potentially existing between the neutron star (NS) and BH mass distributions \citep{Zevin20}.

Meanwhile, the observed higher explosion energy of HNe than the canonical value \citep[$\sim10^{51}$\,erg, see, e.g.][]{Morozova18} can be modelled by the collapse and explosion of massive stars typically with $M_{\rm ZAMS}\gtrsim30-40$\,$M_\odot$ \citep[e.g.,][]{iwamoto98,MazzaliSN1998ef}.
Furthermore there are well known events, Type Ic SN 1998bw \citep{iwamoto98} and LGRB 980425 \citep{Galama98}, which for the first time associated a HN and an LGRB (see also \citet{Cano2017a} for a review).
These facts can naturally lead us to assume that HNe and LGRBs are closely related to the BH formation.

As a possible route to the LGRBs and HNe, \cite{MacFadyen99} proposed the collapsar scenario.
In their scenario, a BH surrounded by a disk is the requisite system, which could be formed in the aftermath of rapidly rotating massive stellar collapse or compact binary coalescence.
Afterward the annihilation of neutrino-antineutrino pairs \citep{Eichler89,Woosley93,Dessart09} or the Blandford-Znajek (BZ) mechanism mediated by strong poloidal magnetic fields threading the BH \citep{Blandford77}, as has been recently demonstrated by \cite{Christie2019dec,Hayashi22} in the context of binary merger, may account for the launch of relativistic GRB jets.
Furthermore the formed massive disk itself could explode, e.g., via viscous heating mechanism \citep{Just22,Fujibayashi23,Coleman24,Fujibayashi:2023oyt} or via magnetohydrodynamics effects associated with the BH spin~\citep{Shibata:2023tho}, resulting in a HN-like explosion energy.

In this study, we explore impacts of progenitor rotation and magnetic fields on the BH formation, focusing on a possible diversity of newly born BH properties, and its subsequent explosion dynamics.
To this end, we conduct 2D axially symmetric magnetohydrodynamics (MHD) simulations for the collapse of a $70$\,$M_\odot$ progenitor star with the two-moment (M1) neutrino transport equation in full general relativity~\citep{KurodaT16}.
A latest report on population of merging BH's properties, such as mass and spin \citep{Abbott23PRX}, also motivated us to direct towards such full-fledged SN simulation beyond BH, as the stellar magnetic field together with neutrino heating/cooling can significantly influence the dynamics at the BH formation and thus potentially the property of proto-BHs.
One remarkable finding in this paper is the emergence of a relativistic jet immediately after BH formation, which powers the explosion up to the HN-like explosion energy, for a progenitor star with a strong poloidal magnetic field and may determine the initial BH evolution.

This paper is organized as follows.
Section~\ref{sec:Method and initial models} starts with a concise summary of our radiation-MHD scheme and the initial setup of the simulation.
The main results are presented in Section~\ref{sec:Results}.
We summarize our results and discussion in Section~\ref{sec:Summary and outlook}.
Throughout this paper, cgs unit is used and Greek indices run from 0 to 3. $c$ and $G$ are the speed of light and gravitational constant, respectively.

\section{Method and initial models}
\label{sec:Method and initial models}
The methodology including neutrino opacities is essentially the same as our former failed supernova simulation in numerical relativity~\citep{KurodaT23BH} other than that we solve MHD equations \citep{KurodaT20}.
We simulate a collapse of a zero-metallicity progenitor model of \cite{Takahashi14}, whose ZAMS mass is $70\,M_\odot$.
At the pre-collapse phase, its iron core mass reaches $\sim2.8$\,$M_\odot$.
In this study we simply use an artificial rotation and magnetic field profile expressed as follows:
\begin{eqnarray}
    u^tu_\phi=\varpi_0^2(\Omega_0-\Omega),
\end{eqnarray}
for the rotational profile $\Omega=\Omega(x)$, where $u^t$ is the time component of the four velocity $u^\mu$, $u_\phi= x u_y$ (in the Cartoon method), $x$ is the distance from rotational axis, and $\varpi_0$ and $\Omega_0$ are parameters denoting the size and angular frequency of a rigidly rotating central cylinder, respectively. Magnetic fields are given through the vector potential $\bf{A}$ as $\bf B=\nabla\times \bf A$, and 
\begin{eqnarray}
    (A_r,A_\theta, A_\phi)=\left(0,0,
    \frac{B_0}{2}\frac{R_0^3}{r^3+R_0^3}r\sin{\theta}
    \right),
\end{eqnarray}
where $B_0$ and $R_0$ represent the magnetic field strength at center and the size of central sphere with uniform magnetic fields, respectively.
$(r,\theta,\phi)$ denote the usual spherical polar coordinates.

Below we present four models: R0B00, which is a non-rotating non-magnetized reference model and has been reported in \cite{KurodaT23BH}, and three rotating models R1B00, R1B11, and R1B12.
These three rotating models employ a fixed initial rotational frequency of $\Omega_0=1$\,rad\,s$^{-1}$.
The two digit after ``B'' denotes the initial strength of central magnetic fields $B_0$ and three different strengths $B_0=0$, $10^{11}$, and $10^{12}$\,G are employed for models R1B00, R1B11, and R1B12, respectively \citep[for discussion about plausible initial magnetic field strength, see,][and Sec.~\ref{sec:Summary and outlook}]{Yoon&Langer12}.
Regarding $\varpi_0$ and $R_0$, we employ often-used values $\varpi_0=R_0=10^8$\,cm, which is also roughly corresponding to the iron core radius \citep{Takahashi14}.

The axisymmetric cylindrical computational domain extends to $1.5\times10^4$\,km along each axis. In the computational domain, nested boxes from 0 to $L_{\rm max}$ refinement levels are embedded and each nested box contains $64\times 64$ cells.
In the present study, we set $L_{\rm max}=11$, 
so that the finest resolution at the center achieves $\sim$115\,m.
The neutrino energy space logarithmically covers from 3 to 300\,MeV with 12 energy bins. We employ the DD2 equation of state (EOS) of \cite{Typel10}, which allows the maximum mass of 2.42\,$M_\odot$ for non-rotating cold NS.

\section{Results}
\label{sec:Results}
%
In this section, we will first report the PNS and BH evolution and will thereafter focus on the emergence of a relativistic jet.

\subsection{Overview of PNS and BH evolution}
\label{sec:Overview of PNS and BH evolution}
%
\begin{figure}
\begin{center}
\includegraphics[angle=-90.,width=1.0\columnwidth]{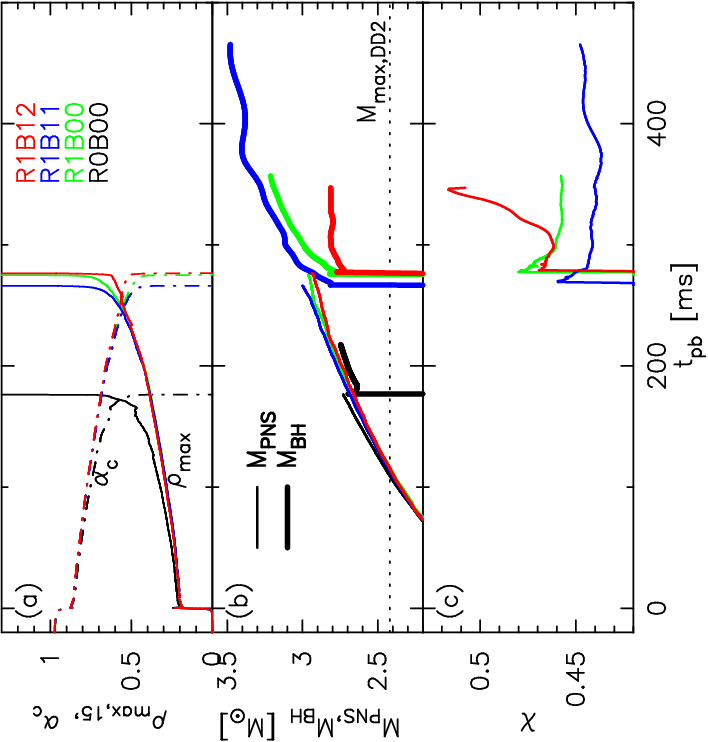}
\caption{{\em Top panel}: evolution of the maximum rest mass density $\rho_{\rm max,15}$ in units of $10^{15}$\,g\,cm$^{-3}$ and the central lapse function $\alpha_{\rm c}$. {\em Middle panel}: evolution of the PNS ($M_{\rm PNS}$: thin) and BH ($M_{\rm BH}$: thick line) masses (see text). The horizontal dotted line marks the maximum mass of cold non-rotating NSs for the DD2 EOS.
{\em Bottom panel}: evolution of the dimensionless BH spin parameter $\chi$.
In all the panels, the color represents the models displayed in the top panel.
\label{fig:Overall}}
\end{center}
\end{figure}
From Fig.~\ref{fig:Overall}, we can grasp overall evolution during the PNS contraction and post-BH formation phase.
From top panel we plot: (a) the maximum density $\rho_{\rm max,15}(\equiv \rho_{\rm max}/10^{15}$\,g\,cm$^{-3}$) (solid lines) and central lapse function $\alpha_{\rm c}$ (dash-dotted), (b) the ADM-based PNS mass $M_{\rm PNS}$ (thin) and BH mass (thick), and (c) the dimensionless BH spin parameter $\chi$ as a function of post-bounce time $t_{\rm pb}$ for all the models denoted by different colors.
In panel (b), we plot the maximum cold NS mass $2.42$\,$M_\odot$ for the DD2 EOS by the horizontal dotted line.
We define the ADM-based PNS mass $M_{\rm PNS}$ by
\begin{eqnarray}
    M_{\rm PNS}&=&\int_{\rho\ge 10^{10}\,\rm{g}\,\rm{cm^{-3}}} \Biggl\{\psi^5\left[ n_\mu n_\nu T^{\mu\nu}+\frac{\tilde A^{ij}\tilde A_{ij}}{16\pi}-\frac{(\hat K+2\Theta)^2}{24\pi} \right] \Biggr.\nonumber \\
    &&\Biggl.-\frac{\tilde \Gamma^{ijk}\tilde \Gamma_{ijk}-(1-\psi)\tilde R}{16\pi}
    \Biggr\}dx^3,
    \label{eq:Mpns}
\end{eqnarray}
where the energy-momentum tensor $T^{\mu\nu}$ takes into account contributions from matter, electromagnetic, and neutrino radiation fields \citep[Eq.~(1) in][]{KurodaT20} and, as for the rest of geometrical variables, we follow notations used in the Z4c formalism \citep{Hilditch13}.
We note that $M_{\rm PNS}$ is used just as a rough measurement of the system as we do not evaluate its value at spatial infinity, but simply by a surface integral (practically its conversion to volume integral Eq.~(\ref{eq:Mpns}) by Gauss's theorem, c.f., \cite{Duez06}) on the isodensity surface at $\rho=10^{10}$\,g\,cm$^{-3}$. 

Black-hole mass $M_{\rm BH}$ is evaluated from the equatorial circumferential radius $C_e$ as~(e.g., \citet{Shibata2016a})
\begin{equation}
    M_{\rm BH}=c^2(4\pi G)^{-1} C_e,
\end{equation}
with $C_e$ in the axisymmetric case being evaluated via
\begin{equation}
    C_e=2\pi\sqrt{\hat g_{\phi\phi}}.
\end{equation}
Here $\hat g_{\mu\nu}$ is the four metric described in the spherical polar coordinates.
The dimensionless BH spin parameter $\chi(=cJ/GM_{\rm BH}^2)$, with $J$ being the angular momentum of the BH, is evaluated via the following relation
\begin{equation}
    \mathcal{A}=8\pi (Gc^{-2}M_{\rm BH})^2\left(1+\sqrt{1-\chi^2}\right),
\end{equation}
where $\mathcal{A}$ is the area of the apparent horizon (AH) \citep{ShibataAH97}.

After core bounce the maximum density and central lapse function in all the models exhibit a nearly monotonically increasing and decreasing trend, respectively, associated with the PNS contraction.
At $t_{\rm pb}\sim175$--$280$\,ms, all the models show the second collapse and form a BH.
The PNS mass at this moment is $\sim2.75$\,$M_\odot$ for R0B00 and $\sim2.9$--$3.0$\,$M_\odot$ for the remaining rotating models.
It is obvious that the rotation delays the onset of BH formation, for the current initial angular velocity, by $\sim100$\,ms.
The reason is that the centrifugal support increases the maximum mass of gravitationally stable rotating PNS by $\sim 20\%$ in comparison to the non-rotating PNS (e.g., \citet{Cook:1993qr}).

Next we explain the evolution of mass and spin parameter of BH for all the models except for R1B12, which will be discussed in the next section.
In the post-BH formation phase, the BH mass evolution tracks a similar trend to that of $M_{\rm PNS}$ for models R0B00, R1B00, and R1B11.
Such a trend is quite reasonable as the BH is first formed from the PNS core and then the residual PNS materials are swallowed by the newly formed BH. 
We indeed confirmed that these models, which do not explode in the simulation time, enter the Bondi-like accretion phase shortly after the BH formation (see \cite{KurodaT23BH} for the case of R0B00).
At this phase the specific angular momentum of the infalling matter is $\sim10^{16}$\,cm$^2$\,s$^{-1}$, which roughly corresponds to $30$--40$\%$ of the value at the innermost stable circular orbit for a $\sim3$\,$M_\odot$ BH.
Regarding the dimensionless spin parameter $\chi$, two rotating models R1B00 and R1B11 initially present a decreasing trend.
It is thereafter followed by a nearly constant phase with $\chi\sim0.44$--$0.46$.
Previous stellar collapse and BH formation simulation by \cite{Fujibayashi23} also reported such an initial trend, i.e. $M_{\rm BH}$ and $\chi$ increases and stagnates, respectively, just after BH formation.
Our rotating models initially possess the total angular momentum of $J\sim4\times10^{49}$\,g\,cm$^2$\,s$^{-1}$ within a sphere with an enclosed rest mass of $\sim3$\,$M_\odot$.
These values give an approximate estimation for the dimensionless spin parameter of proto-BH and it becomes $\sim0.5$, which is in line with the evaluated values of $\chi$ in panel (c).


\subsection{Relativistic jet formation}
\label{sec:Relativistic jet formation}
\begin{figure}
\begin{center}
\includegraphics[angle=-90.,width=\columnwidth]{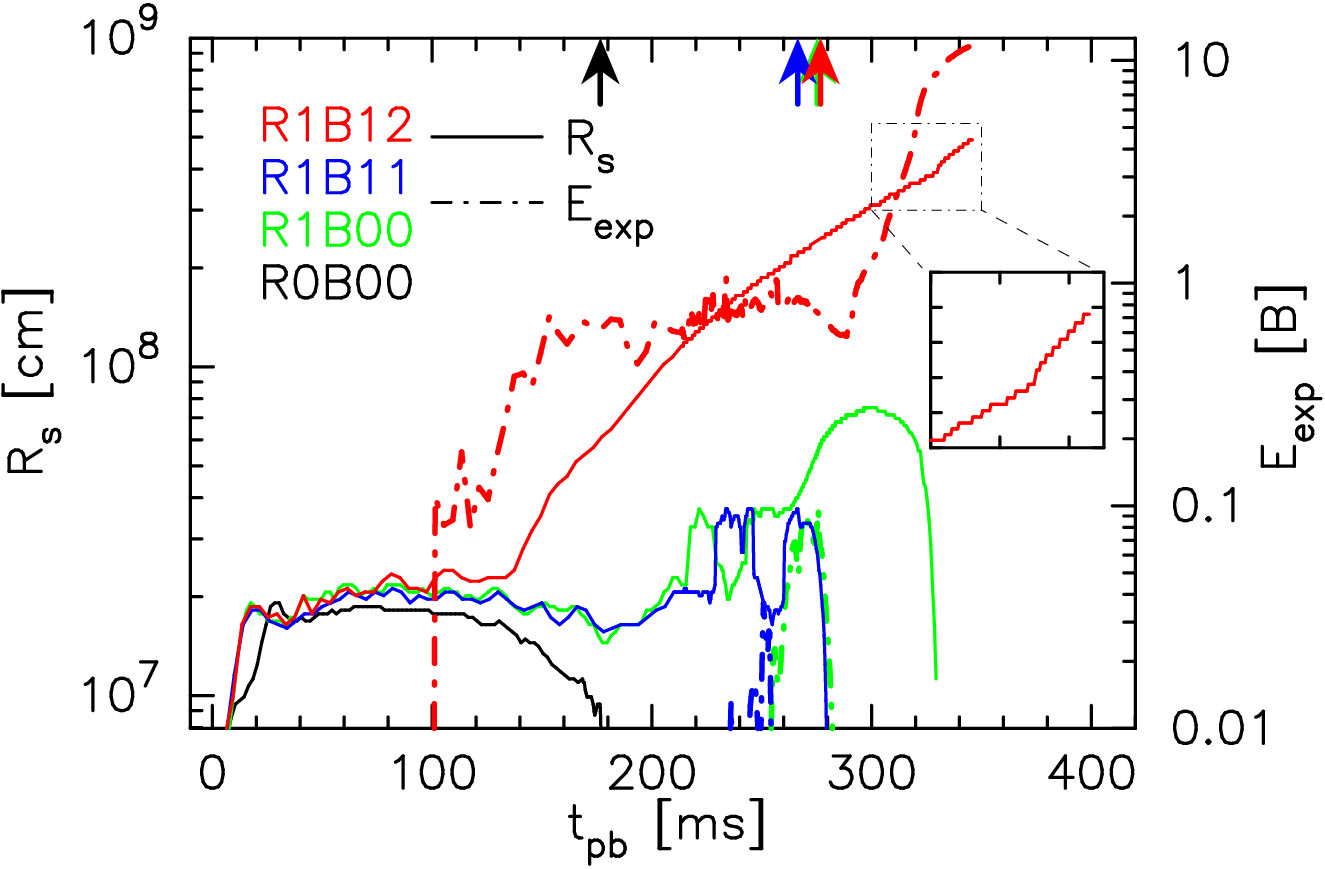}
\caption{Evolution of the maximum shock radius $r_{\rm shock}$ (solid line) and diagnostic explosion energy $E_{\rm exp}$ for all the models.
The BH formation time is marked by the arrow on the top side.
The inset marks when the jet passes the first shock.
\label{fig:RshockEexp}}
\end{center}
\end{figure}
\begin{figure*}
\begin{center}
\includegraphics[angle=-90.,width=\columnwidth]{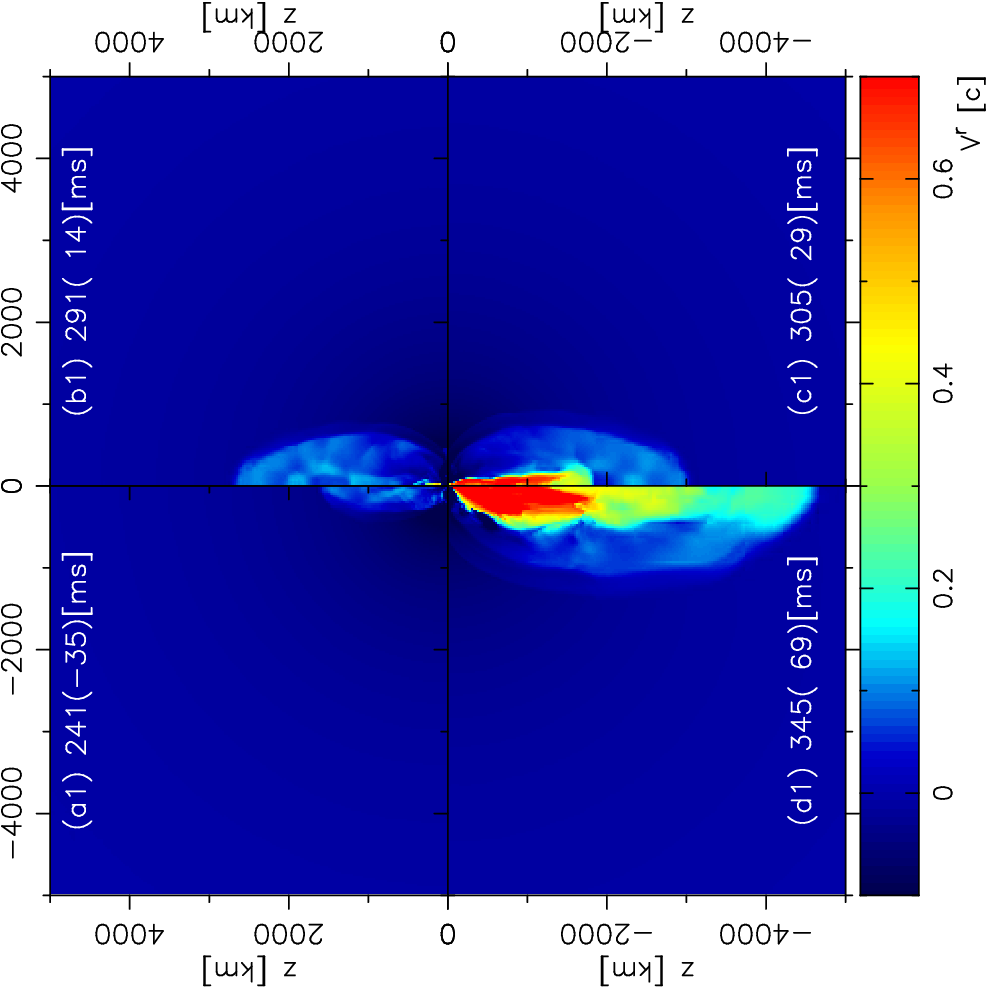}
\includegraphics[angle=-90.,width=\columnwidth]{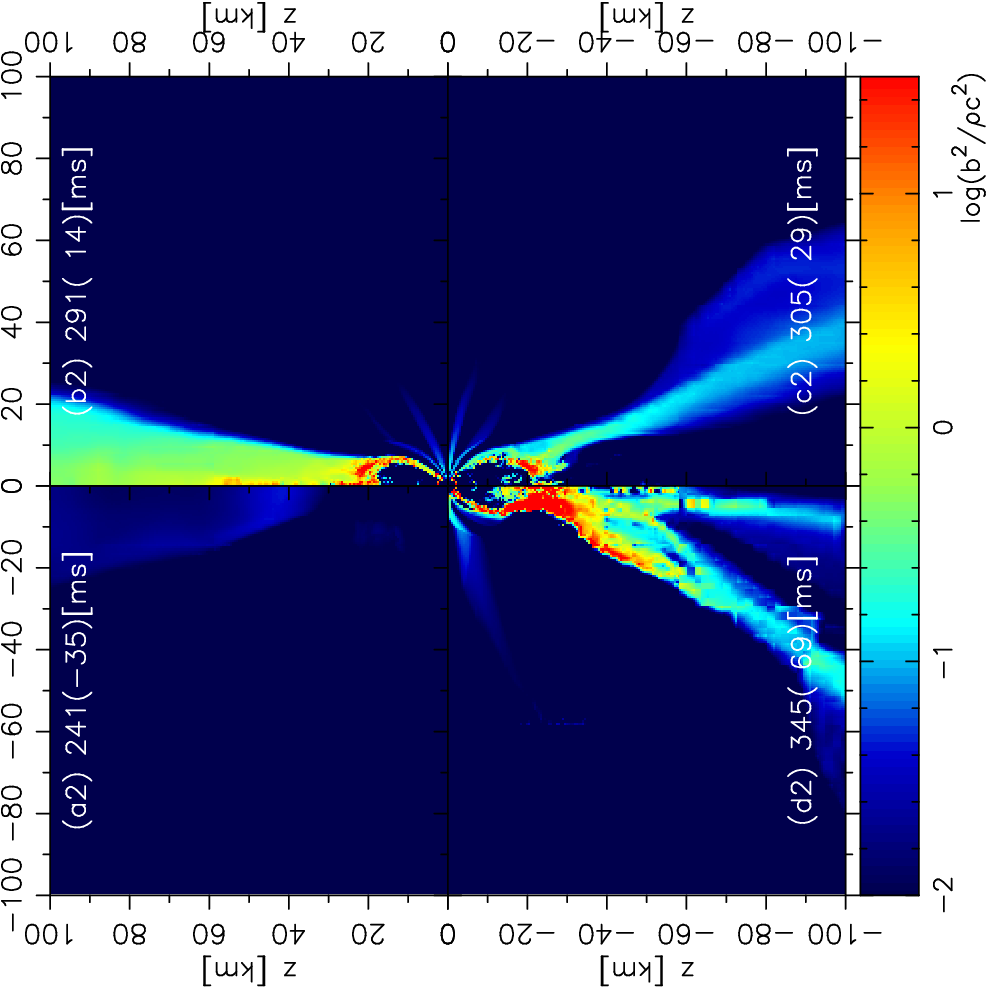}
\caption{Several time snapshot of $v^r$ (left) and $\log (b^2/\rho c^2)$ (right) on the $x$-$z$ plane for R1B12.
Each mini-panel (a1,2)--(d1,2) shows a different post-bounce and post-BH formation time, with the latter being denoted in parentheses.
The right panel focuses on the inner 100\,km region, while a more global region ($\le5000$\,km) is depicted in the left panel.
\label{fig:Beta_Vr}}
\end{center}
\end{figure*}
Fig.~\ref{fig:RshockEexp} presents the maximum shock radius $R_{\rm s}$ (solid line) and diagnostic explosion energy $E_{\rm exp}$ (dash-dotted) for all the models distinguished by color.
The diagnostic explosion energy $E_{\rm exp}$ is evaluated following a definition of \cite{BMuller12a}, but taking into account the additional contribution from magnetic fields.
For reference the BH formation is marked by the arrow on the upper side.

The figure clearly indicates that R1B12 (red solid line) enters the shock expansion phase already at $t_{\rm pb}\sim140$\,ms, which is significantly earlier than the BH formation time $t_{\rm pb}\sim280$\,ms.
The main driving force is the strong magnetic fields, which are amplified along the rotational axis via winding and compression.
The diagnostic explosion energy stays around $E_{\rm exp}\sim0.9$\,B till the BH formation.
Although the shock expansion has initiated significantly earlier than the BH formation, $\sim140$\,ms before, it was not enough to suppress the mass accretion onto the PNS and to hinder the eventual BH formation.
Consequently model R1B12 exhibits a comparable BH formation time to other rotating models (see blue, green, and red arrows).

In the post-BH formation, $E_{\rm exp}$ shows a remarkable increase.
At the end of the simulation it exceeds 10\,B (1\,B=1\,Bethe=$10^{51}$\,erg) and is still increasing exponentially as shown by the red dash-dotted line.
In addition, we can see a kink in the shock evolution (red solid line) at around $t_{\rm pb}\sim330$\,ms.
This is the moment at which the second relativistic jet overtakes the first bipolar shock front.
As a comparison with previous HN-models, \cite{Obergaulinger21} reported $E_{\rm exp}\sim10$\,B at $t_{\rm pb}\sim1$\,s in one of their 3D models due to continuous mass ejection.
One plausible mechanism to explain our high explosion energy and jet formation is the magnetic field amplification associated with the second collapse and the extraction of angular momentum from rotating BH~\citep{Woosley93,Paczynski98}.

Regarding R1B00/B11, they also experience a shock expansion prior to BH formation.
In R1B00, the shock expansion is driven by the neutrino heating, which is confirmed by the ratio of neutrino heating to advection time scale (\cite{Buras06a}, Kuroda et al. in preparation), exceeding unity.
However in these models, the BH formation suddenly suppresses the neutrino irradiation and also the magnetic fields are not so strong to further energize the first shock expansion, resulting in the shock recession as can be seen from green and blue lines in Fig.~\ref{fig:RshockEexp}.

To see how the magnetic fields play their role we depict in Fig.~\ref{fig:Beta_Vr} the radial component of the three velocity $v^r(=u^r/u^t)$ in units of the speed of light $c$ (left panel) and the ratio of magnetic energy density $b^2=b^\mu b_\mu$ to the rest mass energy density $\rho c^2$ in logarithmic scale (right panel).
Here $b^\mu$ is the four vector of magnetic field in the fluid rest-frame.
Each mini-panel (a1,2)--(d1,2) shows a different post-bounce and post-BH formation time ($t_{\rm BH}$), with the latter being denoted in parentheses.
We note that the right panel focuses on the inner 100\,km region, while a more global region ($\le5000$\,km) is shown in the left panel.

From mini-panel (a1) we can observe the first structure of bipolar outflow.
At $t_{\rm pb}(t_{\rm BH})=241(-35)$\,ms, the shock front has already reached to $z\sim1500$\,km.
The shock structure is the commonly observed magneto-rotational explosion (MRE) profile \citep[for recent studies, see, e.g.,][]{Moesta18,KurodaT20,Bugli20,Obergaulinger21} and is driven by the strongly amplified, twisted magnetic fields above the PNS (c.f. $\sim10^{16}$\,G at $z\sim20$\,km).
While the ratio of its energy density to the rest-mass density is still low $\log (b^2/\rho c^2)\lesssim-2$ prior to BH formation (a2), we can observe some regions with the value exceeding unity in the post-BH formation (see red regions in panels (b2)--(d2)).
This is well suited for the formation of a relativistic jet which could launch a gamma-ray burst.
The region with $z\sim20$\,km is the base of the jet, from where the {\it low-}density components are expelled at relativistic speeds ($v_r\sim0.7c$ as seen in panels (c1) and (d1)) by strong magnetic fields.
Here the emergence of the low-density (though not shown, $\rho\sim10^{7\mbox{--}8}$\,g\,cm$^{-3}$) region at $z\sim10$--20\,km is due partly to the BH formation as it suddenly swallows surrounding materials.
Additionally the magneto-centrifugal force expels materials away from the rotational axis.
Indeed we confirmed that the density at a slightly off axis region (e.g., at $z\sim20$\,km and $\theta\sim10^\circ$) exhibits a sharp density increase by about three orders of magnitude compared to those on axis.
Such a feature is also observed in other rotating models R1B00 and R1B11.
Consequently a combination of relatively low density components and strong magnetic fields results in the formation of relativistic jet in model R1B12.
This mechanism is essentially the same as ``the magnetic switch'' model originally proposed by \cite{Wheeler02}.

Here we shortly discuss the feasibility of shock breakout.
The current $70$\,$M_\odot$ progenitor model has a binding energy ahead of the shock ($r\gtrsim5\times10^8$\,cm) of $\sim3.5\times10^{51}$\,ergs.
It indicates that, even if the shock eventually engulfs all these stellar envelopes, the residual is still large and positive $> 6.5$\,B.
Although longer time simulation is essential to assess the final value, it is significantly high and might be able to explain the observed HN branch for very massive progenitor stars \citep{Nomoto06,Tanaka09}.

\begin{figure}
\begin{center}
\includegraphics[angle=-90.,width=\columnwidth]{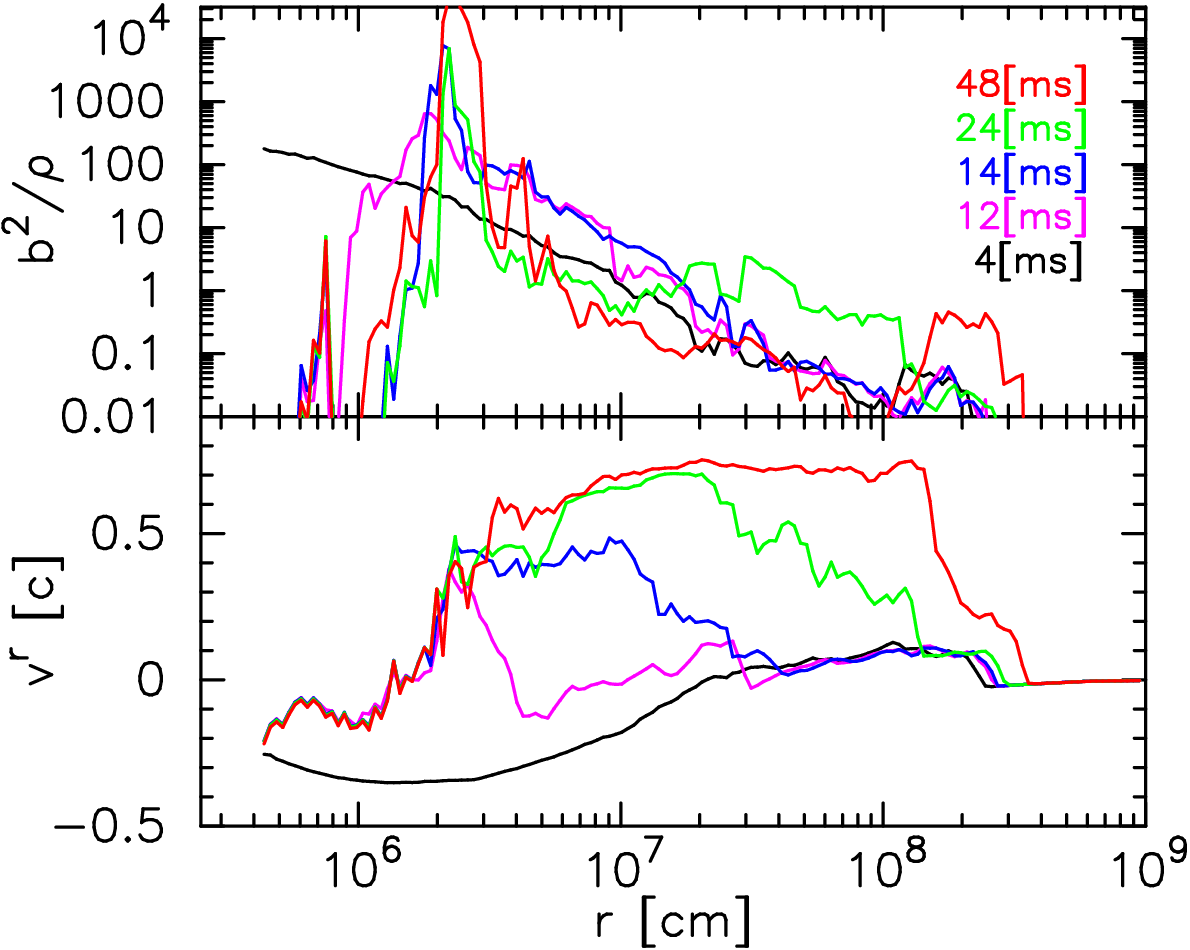}
\caption{Radial profiles of the angle averaged $b^2/\rho c^2$ (top panel) and $v^r$ (bottom) for R1B12 at five different $t_{\rm BH}$ denoted in the top panel.
The angle average is performed for $\theta\le10^\circ$.
\label{fig:JetFormation}}
\end{center}
\end{figure}
To present more clearly that the base of the jet is dominated by strong magnetic fields, we show in Fig.\ref{fig:JetFormation} the radial profile of $b^2/\rho c^2$ (top panel) and $v^r$ (bottom panel) at five different post-BH formation times shown in the upper panel.
Shortly after BH formation ($t_{\rm BH}=4$\,ms) one can find two distinct regions in the velocity profile: outgoing ($r\gtrsim300$\,km) and a nearly free fall region ($r\lesssim100$\,km).
The former corresponds to the first MRE outflow component.
At $t_{\rm BH}=12$\,ms the positive velocity component is gradually appearing at $r\sim30$\,km.
At the same time the ratio $b^2/\rho c^2$ becomes significantly high $\sim10^3$ at $r\sim10$--20\,km as is shown by the magenta line in the top panel, which accelerates the jet to $\sim0.7c$.

After the jet formation, the matter flow in the vicinity of BH is the mass accretion and ejection along the equator and pole, respectively.
And we may attribute the aforementioned BH mass and spin evolution seen in R1B12 (red lines in Fig.~\ref{fig:Overall}) to this characteristic flow channel.
This is because, the suppression of mass infall along the pole simply reduces the total mass infall rate.
Moreover the materials along the pole have less angular momenta than those on the equator, resulting in that their suppression effectively increase $J$ relative to $M_{\rm BH}$ and thus $\chi(=J/M_{\rm BH}^2)$ becomes higher in R1B12.

\section{Summary and outlook}
\label{sec:Summary and outlook}
In this {\it Letter} we have conducted axisymemtric radiation MHD simulations of a core collapse of rotating magnetized $70$\,$M_\odot$ star and its nascent BH evolution in full general relativity.
The current work corresponds to an extension of our previous work~\citep{KurodaT23BH},  which considered only a non-rotating non-magnetized case (model R0B00 in this work).
In addition to R0B00, we reported three new rotating models with different initial magnetic field strengths: non-magnetized (R1B00), moderate (R1B11) and strong (R1B12) one, but in this {\it Letter}, we focused particularly on model R1B12, whose initial magnetic fields are the strongest among the models employed.
Detailed results for the remaining models will be reported in our upcoming paper.

Model R1B12 experienced the so-called MRE at $\sim140$\,ms after core bounce.
The overall feature of the explosion is consistent with previous SN models in which comparably strong initial magnetic fields ($B_0\sim10^{12}$\,G) are employed~\citep[e.g.,][]{Bisnovatyi-Kogan70,LeBlancWilson70,Obergaulinger06,Burrows07,Takiwaki09,Scheidegger10,Moesta18,KurodaT21,Bugli20}.
The bipolar outflow, however, could not completely suppress the mass accretion onto the PNS, and thus, it subsequently collapses to a BH.
Model R1B12 as well as the other rotating models show the BH formation at a similar post-bounce time of $\sim280$\,ms, which is though noticeably later than for the non-rotating case.
Although our rotating models indicated a delayed BH formation, \cite{Rahman2022} reported an opposite trend, as the explosion occurring in a non-rotating model due to its higher neutrino luminosities than those from the counterpart rotating model delays the BH formation.
In the present study, we also found a general feature of stellar rotation on the emergent neutrino profiles (Kuroda~et~al.~in preparation), namely the progenitor rotation basically lowers both neutrino luminosity and mean energy in all flavours, which is consistent with, e.g., \cite{Summa18,Rahman2022}.
In model R1B12, the shock speeds before the BH formation are $\sim0.1c$ and the diagnostic explosion energy accordingly increases to $\sim1$\,B.

After BH formation we observed the emergence of a relativistic jet.
The jet is driven by strong magnetic fields.
For this model for which the magnetic-field strength was already high at the BH formation, we attribute this increase to the density decrease rather than the magnetic field amplification via winding.
Indeed the matter along the pole above the AH is quickly swallowed into the BH.
These two effects reduce the density by about 3--4 orders of magnitude, from $\sim10^{11}$\,g\,cm$^{-3}$ to $\sim10^{7\mbox{--}8}$\,g\,cm$^{-3}$ within a few ms.
Consequently those low-density materials are ejected by strong magnetic fields with $\gg \sqrt{8\pi \rho (v^r)^2}$ at a relativistic speed $\sim0.7c$ inside the cavity of the first bipolar outflow.

Here we shortly touch the kink instability as a possible origin of jet disruption.
Along the jet axis, we found a region ($z\lesssim20$\,km), where the toroidal magnetic fields are three to four orders of magnitude stronger than the poloidal ones.
The region may thus be subject to the kink instability \citep{Lyubarskii99,Begelman98}, which may displace the jet center from the rotational axis and prevent the magnetic field amplification preferentially on the axis \citep{Li00}.
Although such non-axisymmetric effects in the context of SNe are still controversial, e.g., \citet{Moesta14,KurodaT20} reported that the kink instability can weaken the bipolar outflow, \citet{Obergaulinger20} found no significant impacts; the current model R1B12 is worth to be explored in full 3D studies.
We, however, naively expect that, even amidst the kink instability, the jet in full 3D models would not be completely destroyed, but rather just be weakened because the energy injection from the spinning BH should continue for a long timescale (see, e.g., \citet{Shibata:2023tho}).  

Another remarkable finding is its diagnostic explosion energy $E_{\rm exp}$.
It increases rapidly and eventually exceeds $10$\,B at $t_{\rm BH}\sim60$\,ms. 
The value of $>10$\,B is significantly higher than the binding energy of $\sim3.5$\,B possessed by the stellar mantle ahead of the jet and the residual amounts to $>6.5$\,B even if the jet swallows all those stellar mantles.
Although longer time simulations are essential to assess the saturated value, the high explosion energy would be able to explain the observed HN branch at the high-mass end of SN progenitors \citep{Nomoto06,Tanaka09}.
There is, however, a caveat. The BZ mechanism can extract a substantial fraction of rotational kinetic energy of the BH even after the jet launch and continues powering the explosion.
Following \citet{Shibata:2023tho}, we evaluate the extractable rotational kinetic energy for model R1B12 ($M_{\rm BH}\sim3$\,$M_\odot$ and $\chi\sim0.5$) and obtain $\sim100$\,B. The timescale for the extraction is also quite short as $<10$\,s since the magnetic-field strength at the horizon is quite high $\sim 10^{16}$\,G. 
Although the predicted injection energy is still within the upper limit of observed values of LGRBs~\citep{Liang08} and associated HNe, it would provoke a potential problem of too high explosion energy, if the BH mass and spin increase further.
To circumvent the problem, one needs to significantly suppress the subsequent mass and angular momentum accretion onto the BH.
Beside the mass ejection via jet, one possible solution is the early formation of a magnetically arrested disk \citep[MAD:][]{Narayan03}.
In mode R1B12, the mass accretion along equator is indeed weakened compared to other models partly by the strong magnetic field, where we observed the local MAD parameter $\phi_{\rm AH,local}=B^r/\sqrt{\rho v^r c}$ \citep{Hayashi23} reaching $\sim30$--40.

Model R1B12 assumes strong initial magnetic fields by contrast to stellar evolution scenarios \citep[e.g.,][]{Yoon&Langer12} to achieve sufficiently strong magnetic fields soon after bounce and see their visible impacts during feasible simulation time.
In reality, some non-linear magnetic field amplification processes, e.g., magneto-rotational instability \citep{Akiyama03}, which is artificially suppressed in the current 2D models, may operate and potentially amplify the week seed magnetic fields to dynamically relevant strengths.
Therefore we naively expect that the current jet formation model, if performed in 3D, may be reproduced even with weaker initial magnetic fields.
The magnetized disk activities~\citep[e.g.,][in the context of merger]{Christie2019dec,Hayashi22} may also widen a possible window of explodability toward weaker initial magnetic field parameters.
To explore these scenarios, we have to simulate the post-BH formation for the order of several seconds \citep{MacFadyen99,Just22,Fujibayashi23,Fujibayashi:2023oyt,Shibata:2023tho,Coleman24} till outer mantles with high angular momentum and fossil magnetic fields accrete and form the accretion disk around the BH, which is left for future work.

\section*{Acknowledgements}
TK acknowledges the members of the CRA for stimulated discussion and useful advice.
Numerical computations were carried out on Sakura and Raven at Max Planck Computing and Data Facility.
This work was in part supported by Grant-in-Aid for Scientific Research (Nos. 20H00158 and 23H04900) of Japanese MEXT/JSPS.
Numerical computations were carried out on Sakura and Raven clusters at Max Planck Computing and Data Facility.

\section*{Data Availability}
The data underlying this article will be shared on reasonable request to the corresponding author.



\bibliographystyle{mnras}
\bibliography{mybib} 








\bsp	
\label{lastpage}
\end{document}